\documentstyle[prl,aps,graphicx]{revtex}
\def\bra{\langle} \def\ket{\rangle}
\def\<{\langle}
\def\>{\rangle}
\begin{document}
\title{Parallel Quantum Computing in a Single Ensemble Quantum Computer}
\author{Gui Lu Long$^{1,2,3,4}$,
L. Xiao$^{1,2}$}
\address{$^1$ Department of Physics, Tsinghua University, Beijing 100084,
China\\
$^2$ Key Laboratory For Quantum Information and Measurements,
Beijing 100084,China\\
$^3$ Center for Atomic and Molecular Nanosciences, Tsinghua
University, Beijing 100084, China}
\date{\today}
\maketitle

\begin{abstract}
We propose a parallel quantum computing mode for ensemble quantum
computer.  In this mode, some qubits can be in pure states while
other qubits  in mixed states. It enables a single ensemble
quantum computer to perform $``$single-instruction-multi-data"
type of parallel computation. In Grover's algorithm and Shor's
algorithm, parallel quantum computing can provide additional
speedup. In addition, it also makes a fuller use of qubit
resources in an ensemble quantum computer. As a result, some
qubits discarded in the preparation of an effective pure state in
the Schulman-Varizani, and the  Cleve-DiVincenzo algorithms can be
re-utilized.
\end{abstract}
\pacs{03.67.Lx, 03.67.Hk, 89.70.+c}

Quantum computer realization schemes can be classified into
single-quantum-computer type where only a single quantum system is
used, e.g. the trap-ion\cite{trap94}, and ensemble quantum
computer(EQC) type such as the liquid NMR
scheme\cite{cory97,gersh97} and the solid state
scheme\cite{yamamoto02} where many copies of quantum systems are
used. Quantum computer uses superposition of states and possesses
quantum parallelism which provides enormous computing power. It
achieves exponential speedup over existing classical computing
algorithms in prime-factorization\cite{rshor} and simulating
quantum systems\cite{lloyd}. However for some problems the speedup
is not exponential. For instance, Grover's algorithm\cite{grover},
shown optimal\cite{op1}, achieves square-root speedup for unsorted
database search. In some other problems, quantum computer can not
achieve any speedup\cite{divin99}. It is natural to explore
additional speedup by making quantum computers working in
parallel, as in classical computation. By running many identical
quantum computers in parallel, unsorted database search can be
speeded up greatly\cite{grover98,gingrich}. Using Liouville space
computation\cite{mbe}, exponentially fast search can be
achieved\cite{brusch,xiaolong}. The speedup is achieved by using
more resources. EQC is a potential place to exploit this
parallelism because there are many molecules in it. Each molecule
is potentially a single quantum computer, and an EQC is
potentially a collection of that number of quantum computers. At
present, an EQC is used as a single-quantum-computer using
effective pure state technique\cite{cory97,gersh97}, apart from
the lack of projective measurement. Though preparing effective
pure state is tedious, Cleve and DiVincenzo\cite{clevediv96},
Schulman and Varizani\cite{schulmanvar98} have proposed efficient
algorithms to produce a portion of qubits in a pure state and
discard some qubits in the completely mixed states.

In this Letter, we introduce the idea of parallel quantum
computing(PQC) in a single EQC. In the PQC a subset of qubits is
prepared in pure state while the other qubits in mixed state. In
one hand, this enables the $``$single-instruction-multi-data" type
of parallel computation in a single EQC for additional speedup,
example for the Grover and the Shor algorithms. In the other hand,
the PQC uses qubits in mixed state and makes a fuller use of the
qubit resources. For instance those qubits discarded in the
Cleve-DiVincenzo \cite{clevediv96} and the
Schulman-Varizani\cite{schulmanvar98} algorithms can now be
re-used. The PQC is the classical parallel  operation of many
single-quantum-computers.

We introduce notations first.  We call a term in a superposed
state as a component, for instance $|\psi_0\ket$  in
$a|\psi_0\ket+b|\psi_1\ket$; a term in a density matrix a
constituent, for instance $|\psi_0\ket\bra \psi_0|$ in
$p_0|\psi_0\ket\bra
\psi_0|+p_1|\psi_1\ket\bra\psi_1|+p_2|\psi_2\ket\bra\psi_2|$. We
can divide an $n$ number qubits system into two parts, one with
$n_1$ qubits and the other with $n_2$ qubits, and $n_1+n_2=n$. The
state of this  $n$ qubits  system may be represented by
$|j_1,j_2\ket$ where $|j_1\ket$ is the first $n_1$ qubits state
and $|j_2\ket$ is the latter $n_2$ qubits state. We can also
combine the two parts to represent the state as
$|j_{12}\ket\equiv|j_1j_2\ket$. We use interchangeably binary and
decimal representations. For instance a 4 qubits state with
$n_1=n_2=2$ can be represented as
$|01,10\ket=|0110\ket=|1,2\ket=|6\ket$, where the first and third
are in the separated binary and decimal forms, whereas the second
and fourth are in the combined binary and decimal forms
respectively.

We then describe the ensemble measurement which is a
generalization of that used in Liouville space
computation\cite{mbe,jcp}. Assume that an EQC can detect the
transition signal from a single molecule. For a molecule with
$n+1$ qubits, one qubit is used as the ancilla qubit and is
labelled 0. The Hamiltonian of the ancilla qubit is
\begin{eqnarray}
H=\omega_0I_{0z}+\sum_{k\>0}2\pi J_{0k}I_{0z}I_{kz},\label{e1}
\end{eqnarray}
where $J_{0k}$ is the $J$-coupling constant between the ancilla
and the $k$-th qubit.  $I_{jz}$ is the $z$-component of the spin
operator for the $j$-th qubit.   The transition frequency of the
ancilla qubit depends on the state of the remaining $n$ qubits. If
the ancilla qubit transition occurs with the $n$ qubits in state
$|i_1i_2...i_n\ket$, its transition frequency is then
$\omega_0+\sum_{k=1}^n\pi J_{0k}(-1)^{i_k}$. This transition
produces a  peak in the ancilla qubit spectrum. For instance, the
$n$ qubits state $|i_1i_2...i_n\ket=|00...0\ket$ corresponds to
the highest frequency $\omega_0+\sum_{k=1}^n \pi J_{0k}$, and the
state $|i_1i_2...i_n\ket=|11...1\ket$ corresponds to the lowest
frequency $\omega_0-\sum_{k=1}^n\pi J_{0k}$. Thus one can tell the
state of the $n$ qubits $|i_1 i_2...i_n\ket$ by looking at this
sign of the multiplet component. Moreover, the ancilla qubit state
itself is represented by the spectral peak direction. If the
ancilla qubit is in the $|0\ket(|1\ket)$ state before transition,
then the spectral peak is upward(downward). The state in the PQC
can be a superposition of basis states, say
$\sum_{j_2=0}^{N_2-1}c_{j_1,j_2}|j_1,j_2\ket$. In this state the
first $n_1$ qubits are in the $|j_1\ket$ and the latter $n_2$
qubits are in superposed state of the $n_2$-register. When we
measure the ancilla qubit, we will observe only one transition.
The transition frequency is random in one of the frequencies
corresponding to the $n_2$ qubits states in states
$|0\ket$,$\cdots$, $|N_2-1\ket$, because the $n$ qubits state will
collapse into one of $N_2$ basis states $|j_1,j_2\ket=|j_1j_2\ket$
randomly with probability $|c_{j_1,j_2}|^2$. When the superposed
state is transformed into a single basis state,  the transition
frequency will be definite and determined by equation (\ref{e1}).
This ancilla qubit spectrum method will serve as the ensemble
measurement throughout this paper. It can tell the ancilla qubit
state by the peak direction, and the $n$ qubits state by the
transition frequency.

Our quantum computer model is an EQC with $N_1=2^{n_1}$ molecules.
Each molecule can be operated and measured. It has $n+m+1$ qubits.
They are divided into 3 parts: 1 ancilla qubit, a function
register with $m$ qubits, and an argument register with $n$
qubits.  The argument register is further divided into two parts:
one part with $n_1$ qubits called $n_1$-register and another part
with $n_2$ qubits called $n_2$-register, and $n=n_1+n_2$. In
general before a computation, the function register and ancilla
qubit are prepared in the pure state $|0\ket$. The argument
register is in a mixed state with $N_1$ constituent. Each
constituent is characterized by the state of the $n_1$-register.
The $n_2$-register in a given constituent is in a superposed state
of its $N_2=2^{n_2}$ basis states. The density operator of the
ensemble is
\begin{eqnarray}
\rho&=&{1\over
N_1}\sum_{j_1=0}^{N_1-1}\left[\sum_{j_2=0}^{N_2-1}c_{j_1,j_2}|0,j_1,j_2\ket\right]
\left[\sum_{j_2=0}^{N_2-1}c_{j_1,j_2}^{\ast}\bra 0,j_1,j_2|\right]
 , \label{mix4}
\end{eqnarray}
where in $|i,j_1,j_2\ket$, $i$, $j_1$ and $j_2$ are the states for
the function, the $n_1$-  and the $n_2$- registers respectively
and $\sum_{j_2=0}^{N_2-1}|c_{j_1,j_2}|^2=1$. The ancilla qubit
state is not written out explicitly.   In this EQC, there are
$N_1$ constituents and $N_1$ molecules. Each molecule is in a
different state, $\sum_{j_2=0}^{N_2-1}c_{j_1,j_2}|0,j_1,j_2\ket$,
which is a superposition of $N_2$ number of computational basis
states.
 In general, a quantum computation performs
unitary transformations on both the argument and the function
registers. Denoting this transformation as $U_c$, the quantum
computation on state (\ref{mix4}) will be
\begin{eqnarray}
\rho\rightarrow\rho_c&=&U_c\rho U_c^{-1}={1\over 2^{n_1}}
\sum_{j_1=0}^{N_1-1}\left[\sum_{j_2=0}^{N_2-1}c_{j_1,j_2}U_c|0,j_1,j_2\ket\right]
\left[\sum_{j_2=0}^{N_2-1}c_{j_1,j_2}^{\ast}\bra0,j_1,j_2|U_c^{\dagger}\right].
\label{mix4pp}
\end{eqnarray}
An ensemble measurement is then performed to read out the result.

The quantum computation represented in Eq. (\ref{mix4pp}) on the
ensemble (\ref{mix4}) is defined as the parallel quantum
computing. In fact it is $N_1$ quantum computers working in
parallel. The computation instruction $U_c$ is the same for all
molecules, but the databases, numbers represented by different
molecules, are different. Hence, the PQC is the
single-instruction-multi-data type of parallel computation in
classical computation. The state (\ref{mix4}) is  the most general
initial state, and  in most applications, the following simplified
state is sufficient: the $n_1$-register in the complete mixed
state $\sum_{j_1=0}^{N_1-1}(1/N_1)|j_1\ket\bra j_1|$ and the
$n_2$-register in the equally weighted superposed state
$\sum_{j_2=0}^{N_2-1}\sqrt{1/N_2}|j_2\ket$. In this case,
$c_{j_1,j_2}=1/\sqrt{N_2}$ for all possible $j_1$ and $j_2$.

Now we apply  the PQC to the Grover algorithm. Suppose the marked
state is $|j_1^0j_2^0\ket$. Only one qubit is required for the
function register in this algorithm. This qubit is also used as
the ancilla qubit for the ensemble measurement. Preparing the
function register in the $|0\ket$ state, the $n_2$-register in the
equally weighted superposed state, and the $n_1$-register in the
complete mixed state, we have then
\begin{eqnarray}
\rho &=& {1\over N_1}\sum_{j_1=0}^{N_1-1}\left[ \sqrt{1\over N_2}
\sum_{j_2=0}^{N_2-1} |0,j_1, j_2\ket\right]\left[ \sqrt{1\over
N_2} \sum_{j_2=0}^{N_2-1} \bra0,j_1,
j_2|\right].\label{stategrover}
\end{eqnarray}
In this way, we divide the database into $N_1$ sub-databases, each
with $N_2$ items. Apply a zero-failure rate Grover
algorithm\cite{longexact} to the ensemble with $J$ iterations,
where $J-1$ is the integer part of $({\pi\over 2}-\beta)/(2\beta)$
 and is approximately $\pi\sqrt{N_2}/4$ and $\beta=\arcsin{1\over
\sqrt{N_2}}$. In this modified Grover algorithm,  each iteration
consists of four steps :1) apply the query to the whole $n$ qubits
argument register, on condition that the query is satisfied,
rotates the phase of the marked state through angle $\phi=2
\arcsin\left(\sqrt{N_2}\sin{\pi\over (4J+6)}\right)$($\phi$ is
slightly smaller than $\pi$); 2) make a Hadmard transformation on
the $n_2$-register; 3) make a phase rotation through angle $\phi$
on the $|0...0\ket$ basis state of the $n_2$-register; 4) make a
Hadmard transformation on the $n_2$-register again.  If a
sub-database  does not contain the marked state, the above
operation does not produce any observable effect. The constituent
that contains the marked item has its $n_1$-register in state
$|j_1^0\ket$.  The modified Grover algorithm transforms its
$n_2$-register from the equally weighted superposed state into a
single state $|j_2^0\ket$ so that the constituent is in the marked
state $|j_1^0j_2^0\ket$. At the end of the modified Grover
algorithm, one makes a further query and on condition that the
query is satisfied, makes a flip on the function register. The
density matrix becomes $ \rho_f=(1/N_1)|0\ket\bra 0|\sum_{j_1\ne
j_1^0}\left[\sum_{j_2=0}^{N_2-1}\sqrt{1/N_2}|j_1j_2\ket\right]
\left[\sum_{j_2=0}^{N_2-1}\sqrt{1/N_2}\bra j_1j_2|\right]
+(1/N_1)|1\ket\bra 1||j_1^0j_2^0\ket\bra j_1^0j_2^0|$.
 Finally, measuring the ancilla
qubit, one obtains $N_1$ transition peaks in the spectrum, each
from a constituent. For those constituents without the marked
item, each peak is upward and its transition frequency is random
in one of those corresponding to  states $|j_1 0\ket$,$\cdots$,
$|j_1 N_2-1\ket$. The constituent with the marked item is in a
unique state and produces a downward peak with definite frequency
corresponding to the state $|j_1^0j_2^0\ket$.  It finds the
marked state with certainty.

The number of queries is about
$\pi\sqrt{N_2}/4=\pi\sqrt{N/N_1}/4$. This is only $1/\sqrt{N_1}$
of that a standard Grover algorithm requires. This is so because
there are $N_1$ single-quantum-computers searching in parallel,
each in  a reduced database with only $N/N_1=N_2$ items. It
requires $\pi\sqrt{N/N_1}/4$ steps for each single-
quantum-computer to complete the search. In one extreme  $n_1=0$,
there is only a single molecule, the number of query is
$\pi\sqrt{N}/4$, which is just that for the standard Grover
algorithm. On the other extreme, if $n_1=n$, $n_2=0$, the EQC
contains $N=2^n$ molecules in completely mixed state, only a
single query is needed. This is just the Liouville space computer
fetching algorithm proposed recently\cite{xiaolong}.  In Liouville
space computation\cite{mbe}, no superposition of the computational
basis states is used. Each molecule can also be viewed as a
reversible classical computer that can be realized  quantum
mechanics\cite{benioff}, or simply be implemented directly using
classical Turing machine with three tapes\cite{bennett}. If
$n_1=n-1$ and $n_2=1$, the algorithm finds the marked item with
just two queries. Clearly, the speedup is achieved at the expense
of more molecules. The number of queries $N_q$ and the number of
molecules $N_1$ satisfy $N_q^2\times N_1=$constant.

If we fix the number of molecules in an EQC, say at $N_E$. Then in
order each constituent is occupied by at least one molecule, $n_1$
can not be larger than $\log_2 N_E$, otherwise there will be
constituent without any occupying molecules. We assume that the
qubit number $n$ is very large, $N_E\le 2^n$. The maximum value
for $n_1$ is $\log_2N_E$. A natural estimate of the bound is to
set $N_E=N_A$, the Avogadro constant. This sets to $n_1\le79$. In
principle, we can vary $n_1$ from 0 to $\log_2N_E$ so that the
functioning of the EQC changes. When $n_1=0$, all $N_E$ molecules
are in the same pure state, the EQC works as a
single-quantum-computer. Most NMR EQC quantum computation
experiments done so far manage to get this effect using the
effective pure state technique. When $n_1=1$, the ensemble is
divided into two sub-ensembles each with $N_E/2$ molecules. Each
sub-ensemble works as a single quantum computer. The whole
ensemble works as two single-quantum-computers in parallel. When
$n_1=\log_2N_E$, the ensemble works as $N_E$
single-quantum-computers working in parallel.

In the above discussion, a single molecule and an ensemble of many
molecules in pure state are all treated as a
single-quantum-computer. We point here that the EQC can do more by
implementing the parallel operation proposed in
Refs.\cite{grover98,gingrich}. In these work, the Grover algorithm
is run on some $k$  identical quantum computers in parallel. It is
equivalent to repeating the algorithm in a single-quantum-computer
$k$ times. We call this parallel algorithm as repetition parallel
algorithm(RPA). For instance, in Ref.\cite{grover98}, by running
one iteration of Grover's algorithm on $k$ number of identical
quantum computers simultaneously and then measuring these quantum
computers simultaneously, the marked state can be found by picking
out the one most quantum computers point to. Because the marked
state will appear $9k/N$ times in the outcome, whereas any other
state appears $k/N$ times. When $k=O(N\log N)$, the probability
that the marked state occurs more than any other state approaches
unity. In Ref.\cite{gingrich}, $k$ identical quantum computers are
searching in parallel. In each quantum computer, the probability
for finding marked state is amplified. Because there are $k$
quantum computers, by using the majority-vote rule, one needs less
iterations on each quantum computer. The speedup scales as
$O(\sqrt{k})$. The extent of speedup is the same as the PQC
algorithm. But there are several differences between the PQC and
the RPA: 1) in the PQC, the database for each quantum computer is
reduced from $N$ to $N/N_1$, whereas in repetition parallelism,
the database size is always $N$; 2) in the PQC, some $n_1$ qubits
are in mixed state, whereas in the RPA, all qubits are in pure
states. This gives the PQC the advantage to make a fuller use of
qubit resources as we will explain later; 3) the PQC algorithm is
of full success rate whereas the RPA is probabilistic. To overcome
fluctuation, it requires more resource than that in the PQC. For
instance, for single query searching, the PQC algorithm requires
$N$ molecules whereas the algorithm in Ref.\cite{grover98}
requires $O(N\log N)$ molecules.

In reality, some number, say $N_s$, of molecules has to be used as
a logical molecule. A logical molecule can be viewed as the
minimum number of molecules that acts as a
single-quantum-computer. Then a molecule in the preceding
discussion should be understood as a logical molecule. The number
of logical molecules in an EQC is $N_E/N_s$. In practice, an NMR
EQC contains a large number of molecules, say $10^{16}$. Though
with effective pure state technique, the number of molecules
contributing to quantum computation is reduced, there are still
$10^{10}$. This is much more than that needed for a logical
molecule. Thus in ensemble quantum computation with effective pure
state technique, it is possible to see the effect of repetition
parallelism. Indeed, it has been pointed out that in ensemble
quantum computation, unsorted database search can be faster than
Grover algorith\cite{collins} by trading space resources with time
resources, a reflection of the repetition parallelism. In
implementing the PQC, effective pure state technique can also be
used to prepare the $n_2+m+1$ qubits in pure state.

Shor algorithm can also be run in the PQC. The aim is to find the
period $r$ of $a^x$ Mod $N_b$. We need two registers, one argument
register with $n$ qubits where $N_b^2< 2^n< 2N_b^2$, and one
function register with similar size. We divide the argument
register into $n_1$ qubits in the complete mixed state, and $n_2$
qubits in pure state. The ensemble is prepared in state described
by (\ref{stategrover}). We perform $a^x$ Mod $N_b$ and store the
results in the function register. After performing Fourier
transform on only the $n_2$-register, the states in the
$n_2$-register becomes identical in all constituents. By measuring
the $n_2$-register using an ancilla qubit, the period in the
$n_2$-register $N_2/r$ can be found. The speedup is achieved due
to two factors. First, the Fourier transform is done on a smaller
space in the  $n_2$-register, and it requires only $O(n_2^2)$
steps as compared with $O(n^2)$ steps in standard Shor algorithm.
Secondly, there are $N_1$ constituents, therefore there are $N_1$
transitions by a single ensemble measurements. In standard Shor
algorithm, several runs of the algorithm are required. In the PQC,
this can be reduced by a factor of $1/N_1$. We illustrate this in
a simple example with $N_b=15$, $a=7$, $n=8$, $n_1=2$, $n_2=6$.
Shor's algorithm in a single-quantum-computer yields the following
state
$|\psi\ket=(|0\ket+|64\ket+|128\ket+...)(|1\ket+|7\ket+|4\ket+|13\ket)$,
and the period in the argument register is $q/r=64$ where $q=256$.
With the PQC, the resulting state is
$\rho=[(|1\ket+|7\ket+|4\ket+|13\ket)]
([00]+[01]+[10]+[11])|[(|0\ket+|16\ket+|32\ket+|48\ket)]$. Upon
measurement, 4 transitions from the $n_2$-register appear, and
this is equivalent to running the algorithm with 6 qubits 4 times.
 But for the PQC operation of Shor
algorithm, there is a restriction on $n_1$: it should not be
large, otherwise the Fourier transformation in $n_2$ qubits will
not achieve the desired destructive interference.

 The PQC uses mixed
state in general. One can take advantage of this  to make a fuller
use of the qubit resources in EQC. In the Cleve-DiVincenzo
\cite{clevediv96}, and the Schulman-Varizani\cite{schulmanvar98}
algorithms, $O(n)$ qubits are prepared in pure state while some
qubits ($O(\sqrt{n})$ in \cite{clevediv96} and $O(n)$ in
\cite{schulmanvar98}) have to be in the completely mixed state and
be discarded. In the PQC, these qubits can be re-used. This gives
a natural criteria for dividing $n$ into $n_1$ and $n_2$. We can
use these discarded qubits as the $n_1$-register. This increases
considerably the number of qubits usable in an EQC.

This work is supported in part by China National Science
Foundation, the National Fundamental Research Program, Contract
No. 001CB309308 and the Hang-Tian Science foundation. Helpful
discussions with Prof. A. Zeilinger, Drs J. W. Pan, C. Brukner,
Ian Glendinning are gratefully acknowledged.

\end{document}